# Fractional spatiotemporal optical vortices

Shunlin Huang,[1, *] Peng Wang,[1] Yilin Xu,[2] Jun Liu,[1, 2, *] and Ruxin Li,[1, 2]

[1]Zhangjiang Laboratory, Shanghai 201210, China

[2] State Key Laboratory of Ultra-intense laser Science and Technology, Shanghai Institute of Optics and Fine Mechanics, Chinese Academy of Sciences, Shanghai 201800, China

*Corresponding authors: huangshunlin@126.com; jliu@siom.ac.cn

Spatiotemporal optical vortices (STOVs) with spiral phase in the space-time domain, which carry intrinsic transverse orbital angular momentum (OAM), introduce a new degree of freedom to light beams and exhibit unique properties. While integer and fractional spatial vortices have been extensively studied and widely applied, and research on integer STOVs have grown prosperously, fractional STOVs (FSTOVs), classified as STOVs with fractional spiral phases are rarely explored due to the challenges in characterizing rapidly varying spatiotemporal phases. Furthermore, approaches for the rapid recognition of FSTOVs are lacking. Herein, we experimentally and theoretically demonstrate the generation of FSTOVs in the far field. The generation, evolution, and diffraction rules of FSTOVs are revealed. Furthermore, a self-referential method for the rapid recognition of FSTOVs based on the energy ratio between the two end lobes of their diffraction patterns is proposed. This work will promote the development of the theory of light with transverse OAM, and open new opportunities for the applications of STOV, such as STOV-based optical communication and quantum information.

**Introduction**

Light with angular momentum, Optical vortex beams, characterized by their spiral phase, have been extensively studied in the spatial domain for decades. Traditional vortex beams exhibit a spatial phase singularity, resulting in an intensity null at the beam center. Since 1992, it has been established that these beams carry orbital angular momentum (OAM) (*1*). The OAM direction of the traditional spatial vortex beam is typically parallel or antiparallel to the beam propagation direction, which is classified as longitudinal OAM. Conventional vortex beams have boost wide applications from micro- to macro-worlds, including microparticle manipulation (*2-4*), super-resolution microscopy (*5*), and optical communications (*6*).

Recently, a novel type of vortex beam with a spiral phase in the space-time domain, known as the spatiotemporal optical vortex (STOV), was observed in filamentation collapse (*7*), and later generated controllably using linear optical setups (*8, 9*), although the STOV had been predicted in earlier years (*10-13*). The OAM orientation of the STOV is orthogonal to the beam propagation direction, resulting in transverse OAM. This transverse OAM, along with transverse spin angular momentum (*14, 15*), refines the theory of angular momentum of light and opens new possibilities in various fields.

Conventional longitudinal OAM vortex beams typically exhibit a phase winding of $l\,2\pi$ along the azimuthal direction where $l$ is an integer. These are referred to as integer vortex beams. However, vortex beams can also possess fractional spiral phases or fractional topological charges (TC), where $l$ is a fraction. Such beams are termed fractional vortex beams (*16-19*). The fractional vortex beams can be regarded as coherent superposition of a series of integer vortex beams (*19*). Unlike integer vortex beams, which exhibit a donut-shaped intensity profile, fractional vortex beams have an open annular shape with a gap in the intensity profile. Additionally, fractional vortex beams are unstable and tend to split during propagation (*16, 18, 19*). Due to their unique properties, fractional vortex beams have been applied in various fields, including cell sorting and orientation (*20, 21*), optical communication (*22, 23*), and image edge enhancement (*24, 25*).

Research on integer STOVs has progressed rapidly, including studies on their generation, evolution, and diffraction (*8, 9, 26-34*), refraction and reflection (*35*), OAM conservation in nonlinear processes (*36-39*). Furthermore, STOVs with unique properties such as toroidal STOV with a ring shape (*32*) and STOV strings with various STOVs carrying in one wave packet (*40*) have also been developed. However, the generation and properties of fractional STOVs (FSTOVs) remain largely unexplored. Moreover, the applicability of the general formula for integer STOVs to FSTOVs generated using 4*f* pulse shaper has not been verified.

Here, we theoretically and experimentally analyze the generation, propagation, and diffraction of FSTOVs with different fractional spiral phases. The wave packet structure of the FSTOV can be clearly demonstrated from its diffraction pattern. The energy of the new lobe generated in the diffraction pattern increases with the fractional TC, allowing for convenient recognition of the fractional TC. Furthermore, we show that the general formula for integer STOVs is not suitable for describing FSTOVs generated using a 4*f* pulse shaper.

**Results**

The schematic of the experimental setup for the generation and recognition of the FSTOVs is shown in Fig. 1. Input pulses are from a mode-locked Ti:sapphire laser. As shown in the top part of Fig. 1, a 4*f* pulse shaper with a spatial light modulator (SLM) at the Fourier plane and a spherical lens ($f = 1$ m) are applied to generate FSTOVs (FSTOVs generator). Then the diffraction patterns of the FSTOVs are captured using a FSTOV detector which is similar to the generator with the SLM replaced by a charge-coupled device (CCD) camera, as depicted in the bottom part of Fig.1. The detector, placed in the focal plane of the spherical lens, functions as a two-dimensional spectrometer, effectively serving as a STOV spectrometer. The parameters of the grating (1200 grooves/mm) and cylindrical lens ($f = 0.3$ m) used in the generator and detector are the same.

The general formula for the electric field of an $l$-th order STOV at a plane $z = 0$ is given by (8)

$$E(x, y, t) = [t/t_0 + i\,\text{sgn}(l)\,y/y_0]^{|l|} E_0(x, y, t), \tag{1}$$

where $E_0$ is a Gaussian pulse in both space and time domains, and $t_0$ and $y_0$ are the temporal and spatial scale widths, respectively. However, we find that this formula is unsuitable for describing FSTOVs generated using a 4$f$ pulse shaper. For instance, the intensity profile of the STOV depicted by Eq. (1) is uniformly distributed in the $y$-$t$ plane and contains only one phase singularity at the center. Moreover, the spectral intensity distribution of FSTOVs obtained by Fourier transform of Eq. (1) does not match the measured results, which are shown in Supplement 1.

To address this, we start from the paraxial wave equation for a uniform isotropic medium, the Helmholtz equation:

$$\nabla^2 E + k^2 E = 0, \tag{2}$$

where $\nabla^2$ is the Laplacian, and $k$ is the wave number. A solution of the equation in the spatial frequency domain is

$$E(f_x, f_y, z) = E(f_x, f_y, z = 0) \cdot H(f_x, f_y), \tag{3}$$

where $H(f_x, f_y) = \exp[ikz - i\pi\lambda z(f_x^2 + f_y^2)]$, $f_x$ and $f_y$ are the spatial frequencies in the $x$- and y-directions, respectively. This solution allows us to calculate the laser pulse fields along the propagation path.

For simplicity, the fields of the input laser pulses are described as Gaussian pulses, which have Gaussian spatial and spectral profiles. The field can be expressed as

$$E_1(x_1, y_1, \omega) = \exp[-(x_1^2 + y_1^2)/a^2]\exp[-(\omega - \omega_c)^2/b^2], \qquad (4)$$

where $a$ and $b$ are the waist radii of the Gaussian profiles in the spatial and spectral domains, respectively, $\omega_c = 2\pi c/\lambda_c$ is the central frequency, $\lambda_c$ is the central wavelength, and $c$ is the vacuum light velocity. This filed is then injected into the 4$f$ pulse shaper to generate FSTOVs. The methods for the simulation of the generation, propagation, and diffraction of the FSTOVs are similar to those used in our previous works (*26, 31*), which are described in Supplement 1.

Fractional spiral phase patterns with $\varphi = 2\pi L$ are loaded onto the SLM to modulate the input pulses to generate FSTOV, where $L$ is a fraction. These fractional spiral phase pattern have $L$ 2π phase winding, corresponding to a TC of $L$. The simulated results of the FSTOVs generated using fractional spiral phases within the ranges of 0-1 and 1-2 are presented in Fig. 2(A) and (B), respectively. In both Fig. 2(A) and (B), rows 1 and 2 show the three-dimensional (3D) iso-intensity profiles of the FSTOVs in the near and far fields, respectively; rows 3 and 4 show the intensity profiles and phase distributions of the corresponding FSTOVs in the far field, respectively.

For FSTOVs with TCs in the 0-1 range, the near-field evolution reveals a gradual transformation from a single solid wave packet into a two-lobe structure as $L$ increase. The splitting of the wave packet is shown clearly in row 1. In the far field, when the TC deviates significantly from the integer values of 0 and 1

(e.g., $L = 0.6$), the wave packet exhibits a distinct opening (or gap). When the TC approaches an integer, a wave packet with a donut-shaped profile is generated.

It is worth noting that this behavior differs from that of conventional spatial fractional vortices, in which the size of the radial opening initially increases with the TC, reaches its maximum at half-integer values, and then decreases as the TC continues to increase (*41*).

Furthermore, as can be observed from rows 2 and 3, as the fractional TC increases, new structures of the FSTOV emerge at the top portion of the wave packet. These structures deviate from the center of the wave packet due to the off-center phase distribution depicted in row 4. Specifically, for FSTOVs with small TCs, such as $L = 0.2$, the phase singularities are not located at the center but rather at the top part of the wave packet. Note that for negative FSTOVs, the phase singularities are situated at the bottom part. As the TC increases, the phase singularity gradually moves toward the center of the wave packet and ultimately reaches the center when the TC becomes an integer.

Interestingly, although the spiral phase in the Fourier plane used to generate the FSTOV is fractional, the phase of the FSTOV in the far field exhibits a $2\pi$ phase winding within a small region around the singularity. However, because the phase singularity is off-center, the FSTOV does not display a $2\pi$ phase winding at the center of the wave packet. This indicates that the FSTOV has not yet evolved into integer STOV.

For FSTOVs with topological charges (TCs) in the 1-2 range, the wave packet undergoes distinct structural transformations. In the near field, as shown in row 1 of Fig. 2(B), the wave packet evolves from a two-lobe structure to a three-lobe structure. In the far field, as depicted in row 2, the wave packet transitions from a donut shape to a two-hole structure. Similar to FSTOVs with TCs in the

0-1 range, these FSTOVs also exhibit off-center phase distributions, as illustrated in row 4. This off-center phase distribution leads to the emergence of new structures at the top portions of the wave packets.

Notably, the position of the spiral phase corresponding to the integer part of the TC remains relatively unchanged as the TC varies and is consistently located at the bottom part of the wave packet. In contrast, the fractional part of the phase is situated at the top part of the wave packet and gradually moves closer to the center as the TC increases.

Without loss of generality, the evolutions of FSTOVs with $L = 0.5$ and 1.5 are presented as examples in Fig. 3(A) and (B), respectively. In both Fig. 3(A) and (B), rows 1, 2, and 3 show the 3D-isointensity profiles, intensity profiles, and phase patterns of the FSTOV at different positions after the spherical lens, respectively. Here, $z = 0f$ and $z = 1f$ represent the near field and far field, respectively. For the FSTOV with $L = 0.5$ and 1.5, the structures exhibit clear openings, as shown in rows 1. During propagation, the size of the wave packet gradually decreases, reaching its minimum at the focal plane. Furthermore, the FSTOVs at all positions exhibit off-center phase distributions, with the phase singularities located at the top portions of the phase patterns. Additionally, FSTOV exhibits an asymmetric mode evolution before and after the focal plane of the spherical lens, while its chirality remains unchanged after passing through the focus.

To experimentally verify that the generated STOV is an FSTOV, we analyze the wave packet structures of the FSTOVs produced using the 4*f* pulse shaper. While detection methods previously developed for integer STOVs (*8, 9*) are effective, they may not be suitable for the rapid detection of FSTOVs due to several challenges: the non-uniform intensity distribution of the wave packet, the rapid phase variations in both the spatial and temporal domains, and the unique phase distribution of FSTOVs, which includes off-center phase singularities and a 2π phase winding near the phase singularity of the fractional part. We demonstrate that the diffraction pattern of the FSTOV in the space domain serves as a mapping of the wave packet structure in the spectral-spatial domain. Moreover, the TCs of the FSTOVs can be easily obtained by analyzing their diffraction patterns.

The diffraction patterns of the FSTOVs with TCs in the ranges of 0-1 and 1-2 are shown in Fig. 4 (A) and (B), respectively. In both Fig. 4(A) and (B), row 1 shows the simulated diffraction patterns, while row 2 shows the corresponding experimental results. It has been demonstrated that for STOVs with integer TCs, such as TC = 1 and 2, their diffraction patterns exhibit 2-lobe and 3-lobe structures, respectively. Additionally, these diffraction patterns feature two energetic end lobes with nearly equal energy, resulting in symmetric structures (*31*). Interestingly, we observe that FSTOVs with TCs in the ranges of 0-1 and 1-2 also have 2-lobe and 3-lobe structures, respectively, as illustrated in Fig. 4. However, unlike integer STOVs, the two end lobes of the FSTOV diffraction patterns have different energies. As the fractional part of the phase increases, an upper-end lobe emerges and grows (for FSTOV with opposite TC, it is on the opposite end), and energy is transferred from the lower lobes to the upper-end lobe. When the TC becomes an integer (e.g., 1 or 2), the upper-end lobe and the bottom-end lobe attain nearly equal energy.

These characteristics allow the determination of the TCs of FSTOVs. For instance, by calculating the energy ratio between the upper-end and bottom-end lobes, a corresponding TC can be derived. The simulated and measured energy ratios for the two series of FSTOVs are shown in Fig. 4(C) and (D), respectively. The blue dashed and olive solid curves represent the Gaussian fitting curves for the simulated and measured results, respectively. The experimental results align well with the calculated results. The fitted energy ratio curves exhibit a Gaussian shape, with a sharp increase in the middle region, primarily due to the Gaussian spectral distribution of the input pulses used to generate the FSTOVs.

It is worth noting that other algorithms can also be employed. For instance, the fractional TCs can be calculated using the energy ratio between the upper-end lobe and the entire diffraction pattern. Since the denominator of the energy ratio is derived from the same diffraction pattern, this kind of detection method is self-referential and does not require additional reference pulses.

The unique diffraction pattern of FSTOVs primarily arises from their distinctive wave packet structure, characterized by a transverse (OAM)-affected spectral intensity distribution (TOASID). The spectral intensity distributions of FSTOVs with topological charges (TCs) in the 1-2 range in the space-spectrum domain are displayed in row 1 of Fig. 5. For comparison, row 2 shows the spatial-spectral intensity distributions of the corresponding integer STOVs calculated using Eq. (1).

As illustrated in row 1, the emergence of an end lobe becomes evident, and its energy increases with the TC. In contrast, row 2 reveals that the newly generated lobes appear between the two end lobes, leading to diffraction patterns that differ significantly from those shown in Fig. 4. Furthermore, the discrepancies between rows 1 and 2 demonstrate that the general formula Eq. (1), which describes integer STOVs, is not applicable to FSTOVs generated using the 4$f$ pulse shaper.

**The transverse OAM densities of FSTOVs**

The transverse OAM (TOAM) of the STOV is currently a topic of debate. In this work, we follow the calculation methods outlined in References (*9, 32, 37*). The linear momentum density of a scalar field $E$ can be expressed as (*9, 32, 37*)

$$\vec{p} = Im(E^* \nabla E), \tag{5}$$

where *Im* denotes the imaginary part, and * indicates the complex conjugate. Then the OAM density $\vec{j}$ can be calculated as

$$\vec{j} = \vec{r} \times \vec{p} = \vec{r} \times (I\nabla\varphi), \qquad (6)$$

where *I* is the intensity distribution, $\varphi$ is the phase distribution, and $\vec{r}$ is the position vector. The TOAM per photon within a volume can be calculated as

$$\frac{TOAM}{photon} = \frac{\hbar \int \vec{r} \times \vec{p}\, dV}{\int E^* E\, dV} = \frac{\hbar \int \vec{r} \times (I\nabla\varphi)\, dV}{\int I\, dV}. \qquad (7)$$

It can be seen that the *TOAM/photon* is associated with the phase distribution and intensity distribution.

The TOAM per photon for the FSTOVs with TCs of *l* = 0.5, 1, and 1.5 are 1.06$\hbar$, 2.81$\hbar$, and 3.48$\hbar$, respectively. The quantum numbers of the TOAM per photon are larger than the corresponding topological charges due to the asymmetric intensity distribution of the FSTOVs generated by the 4*f* pulse shaper (*32, 38*). Their corresponding OAM densities are shown in Fig. 6. For these three FSTOVs, the OAM densities wind around the phase singularities, forming vortices. As mentioned in Fig. 2, the FSTOVs have off-center phase singularities, causing the OAM density vortices to also deviate from the center of the wave packet.

**Discussion**

In summary, we have experimentally and theoretically analyzed the generation, propagation, and diffraction of FSTOVs with different TCs. Furthermore, we have verified the applicability of the general formula Eq. (1), commonly used to describe integer STOVs, and demonstrated that it is not suitable for characterizing FSTOVs generated using a 4*f* pulse shaper. Our work reveals that FSTOVs can be generated in the far field, which holds significant potential for practical applications. We have also uncovered the evolution rules of FSTOVs and proposed a self-referential method for the rapid recognition of TCs by analyzing the energy ratios derived from diffraction patterns. This work enhances the understanding of the physical properties of STOVs, highlighting that STOVs, like conventional spatial vortex beams, can possess both integer and fractional TCs. Furthermore, FSTOVs hold unique properties compared to spatial fractional vortices, which could advance their applications in various fields, such as STOV-based optical communication, light-matter interaction, image processing (*42*), and quantum information.

**Materials and Methods**

**Experiments**

Laser pulses from a Ti:sapphire mode-locked laser with a central wavelength at 800 nm are injected into the 4*f* pulse shaper to generate FSTOVs. The spectral bandwidth of the input pulses is tailored to about 5 nm. The 4*f* pulse shaper comprises a ruled reflective grating (1200 grooves/mm, blaze wavelength of 750 nm), a cylindrical lens ($f$ = 300 mm), and a two-dimensional reflective SLM (Meadowlark Optics, P1920-HDMI). The distances between the grating, cylindrical lens and the SLM are all set to 300 mm. In order to generate FSTOVs, fractional spiral phase patterns are load onto the reflective SLM, which is positioned at the Fourier plane of the 4*f* pulse shaper, to modulate the input pulses. The modulated pulses exiting the pulse shaper are reflected by a beam splitter and then pass through a spherical lens ($f$ = 1 m). The FSTOVs are generated in the focal plane of the spherical lens. The generated FSTOVs are diffracted by a second grating placed in the focal plane of the spherical lens and collimated by a second cylindrical lens. The diffraction patterns of the FSTOVs are captured by a CCD camera placed in the focal plane of the second cylindrical lens. The parameters of the second grating and cylindrical lens are identical to those used in the pulse shaper. The distance between the second grating and the second cylindrical lens is also set to 300 mm. The detection setup functions as a STOV spectrometer, capable of measuring the TCs of both integer STOVs and FSTOVs.

**Acknowledgments**

**Funding:** This work is supported by
National Natural Science Foundation of China (nos. 12374320)
Natural Science Foundation of Shanghai (nos. 23ZR1471700).

**Competing interests:** The authors declare no conflicts of interest.

**Data availability:** Data underlying the results presented in this paper are not publicly available at this time but may be obtained from the authors upon reasonable request.

**REFERENCES**

1. L. Allen, M. W. Beijersbergen, R. J. C. Spreeuw, J. P. Woerdman, Orbital angular momentum of light and the transformation of Laguerre-Gaussian laser modes. *Phys. Rev. A* **45**, 8185-8189 (1992).
2. H. He, M. E. J. Friese, N. R. Heckenberg, H. Rubinsztein-Dunlop, Direct observation of transfer of angular momentum to absorptive particles from a laser beam with a phase singularity. *Phys. Rev. Lett.* **75**, 826-829 (1995).
3. L. Paterson, M. P. MacDonald, J. Arlt, W. Sibbett, P. E. Bryant, K. Dholakia, Controlled rotation of optically trapped microscopic particles. *Science* **292**, 912-914 (2001).
4. S. Huang, P. Wang, X. Shen, J. Liu, R. Li, Multicolor concentric ultrafast vortex beams with controllable orbital angular momentum. *Appl. Phys. Lett.* **120**, 061102 (2022).
5. L. Yan, P. Gregg, E. Karimi, A. Rubano, L. Marrucci, R. Boyd, S. Ramachandran, Q-plate enabled spectrally diverse orbital-angular-momentum conversion for stimulated emission depletion microscopy. *Optica* **2**, 900-903 (2015).
6. J. Wang, J.-Y. Yang, I. M. Fazal, N. Ahmed, Y. Yan, H. Huang, Y. Ren, Y. Yue, S. Dolinar, M. Tur, A. E. Willner, Terabit free-space data transmission employing orbital angular momentum multiplexing. *Nat. Photonics* **6**, 488-496 (2012).
7. N. Jhajj, I. Larkin, E. W. Rosenthal, S. Zahedpour, J. K. Wahlstrand, H. M. Milchberg, Spatiotemporal optical vortices. *Phys. Rev. X* **6**, 031037 (2016).
8. S. W. Hancock, S. Zahedpour, A. Goffin, H. M. Milchberg, Free-space propagation of spatiotemporal optical vortices. *Optica* **6**, 1547-1553 (2019).
9. A. Chong, C. Wan, J. Chen, Q. Zhan, Generation of spatiotemporal optical vortices with controllable transverse orbital angular momentum. *Nat. Photonics* **14**, 350-354 (2020).

10. A. P. Sukhorukov, V. V. Yangirova, Spatio-temporal vortices: properties, generation and recording. *Proc. SPIE* **5949**, 594906 (2005).
11. N. Dror, B. A. Malomed, Symmetric and asymmetric solitons and vortices in linearly coupled two-dimensional waveguides with the cubic-quintic nonlinearity. *Physica D* **240**, 526-541 (2011).
12. K. Y. Bliokh, F. Nori, Spatiotemporal vortex beams and angular momentum. *Phys. Rev. A* **86**, 033824 (2012).
13. J. Qiu, B. Shen, X. Zhang, Z. Bu, L. Yi, L. Zhang, Z. Xu, Vortex beam of tilted orbital angular momentum generated from grating. *Plasma Phys. Control. Fusion* **61**, 105001 (2019).
14. A. Aiello, N. Lindlein, C. Marquardt, G. Leuchs, Transverse angular momentum and geometric spin Hall effect of light. *Phys. Rev. Lett.* **103**, 100401 (2009).
15. A. Aiello, P. Banzer, M. Neugebauer, G. Leuchs, From transverse angular momentum to photonic wheels. *Nat. Photonics* **9**, 789-795 (2015).
16. M. W. Beijersbergen, R. P. C. Coerwinkel, M. Kristensen, J. P. Woerdman, Helical-wavefront laser beams produced with a spiral phaseplate. *Opt. Commun.* **112**, 321-327 (1994).
17. I. V. Basistiy, M. S. Soskin, M. V. Vasnetsov, Optical wavefront dislocations and their properties. *Opt. Commun.* **119**, 604-612 (1995).
18. M. V. Vasnetsov, I. V. Basistiy, M. S. Soskin, Free-space evolution of monochromatic mixed screw-edge wavefront dislocations. *Proc.SPIE* **3487**, 29-33 (1998).
19. M. V. Berry, Optical vortices evolving from helicoidal integer and fractional phase steps. *J. Opt. A-Pure Appl. Opt.* **6**, 259 (2004).
20. R. Dasgupta, S. Ahlawat, R. S. Verma, P. K. Gupta, Optical orientation and rotation of trapped red blood cells with Laguerre-Gaussian mode. *Opt. Express* **19**, 7680-7688 (2011).
21. S. H. Tao, X. C. Yuan, J. Lin, X. Peng, H. B. Niu, Fractional optical vortex beam induced rotation of particles. *Opt. Express* **13**, 7726-7731 (2005).
22. Z. Xu, C. Gui, S. Li, J. Zhou, J. Wang, Fractional Orbital Angular Momentum (OAM) Free-Space Optical Communications with Atmospheric Turbulence Assisted by MIMO Equalization. *Adv. Photon. Commun.*, JT3A.1 (2014).
23. G. Zhu, Z. Bai, J. Chen, C. Huang, L. Wu, C. Fu, Y. Wang, Ultra-dense perfect optical orbital angular momentum multiplexed holography. *Opt. Express* **29**, 28452-28460 (2021).
24. G. Situ, G. Pedrini, W. Osten, Spiral phase filtering and orientation-selective edge detection/enhancement. *J. Opt. Soc. Am. A* **26**, 1788-1797 (2009).
25. M. K. Sharma, J. Joseph, P. Senthilkumaran, Fractional vortex dipole phase filter. *Appl. Phys. B-Lasers Opt.* **117**, 325-332 (2014).
26. S. Huang, P. Wang, X. Shen, J. Liu, Properties of the generation and propagation of spatiotemporal optical vortices. *Opt. Express* **29**, 26995-27003 (2021).

27. S. W. Hancock, S. Zahedpour, H. M. Milchberg, Mode structure and orbital angular momentum of spatiotemporal optical vortex pulses. *Phys. Rev. Lett.* **127**, 193901 (2021).
28. Q. Cao, J. Chen, K. Lu, C. Wan, A. Chong, Q. Zhan, Sculpturing spatiotemporal wavepackets with chirped pulses. *Photonics Res.* **9**, 2261-2264 (2021).
29. J. Chen, C. Wan, A. Chong, Q. Zhan, Experimental demonstration of cylindrical vector spatiotemporal optical vortex. *Nanophotonics* **10**, 4489-4495 (2021).
30. J. Chen, C. Wan, A. Chong, Q. Zhan, Subwavelength focusing of a spatio-temporal wave packet with transverse orbital angular momentum. *Opt. Express* **28**, 18472-18478 (2020).
31. S. Huang, P. Wang, X. Shen, J. Liu, R. Li, Diffraction properties of light with transverse orbital angular momentum. *Optica* **9**, 469-472 (2022).
32. C. Wan, Q. Cao, J. Chen, A. Chong, Q. Zhan, Toroidal vortices of light. *Nat. Photonics* **16**, 519-522 (2022).
33. W. Chen, W. Zhang, Y. Liu, F.-C. Meng, J. M. Dudley, Y.-Q. Lu, Time diffraction-free transverse orbital angular momentum beams. *Nat. Commun.* **13**, 4021 (2022).
34. L. Gu, Q. Cao, Q. Zhan, Spatiotemporal optical vortex wavepackets with phase singularities embedded in multiple domains [Invited]. *Chin. Opt. Lett.* **21**, 080003 (2023).
35. M. Mazanov, D. Sugic, M. A. Alonso, F. Nori, K. Y. Bliokh, Transverse shifts and time delays of spatiotemporal vortex pulses reflected and refracted at a planar interface. *Nanophotonics* **11**, 737-744 (2022).
36. S. W. Hancock, S. Zahedpour, H. M. Milchberg, Second-harmonic generation of spatiotemporal optical vortices and conservation of orbital angular momentum. *Optica* **8**, 594-597 (2021).
37. G. Gui, N. J. Brooks, H. C. Kapteyn, M. M. Murnane, C.-T. Liao, Second-harmonic generation and the conservation of spatiotemporal orbital angular momentum of light. *Nat. Photonics* **15**, 608-613 (2021).
38. K. Y. Bliokh, Spatiotemporal vortex pulses: angular momenta and spin-orbit interaction. *Phys. Rev. Lett.* **126**, 243601 (2021).
39. M. A. Porras, Transverse orbital angular momentum of spatiotemporal optical vortices. *Prog. Electromagn. Res.* **177**, 95-105 (2023).
40. S. Huang, Z. Li, J. Li, N. Zhang, X. Lu, K. Dorfman, J. Liu, J. Yao, Spatiotemporal vortex strings. *Sci. Adv.* **10**, eadn6206 (2024).
41. W. M. Lee, X. C. Yuan, K. Dholakia, Experimental observation of optical vortex evolution in a Gaussian beam with an embedded fractional phase step. *Opt. Commun.* **239**, 129-135 (2004).
42. J. Huang, J. Zhang, T. Zhu, Z. Ruan, Spatiotemporal differentiators generating optical vortices with transverse orbital angular momentum and detecting sharp change of pulse envelope. *Laser Photon. Rev.* **16**, 2100357 (2022).

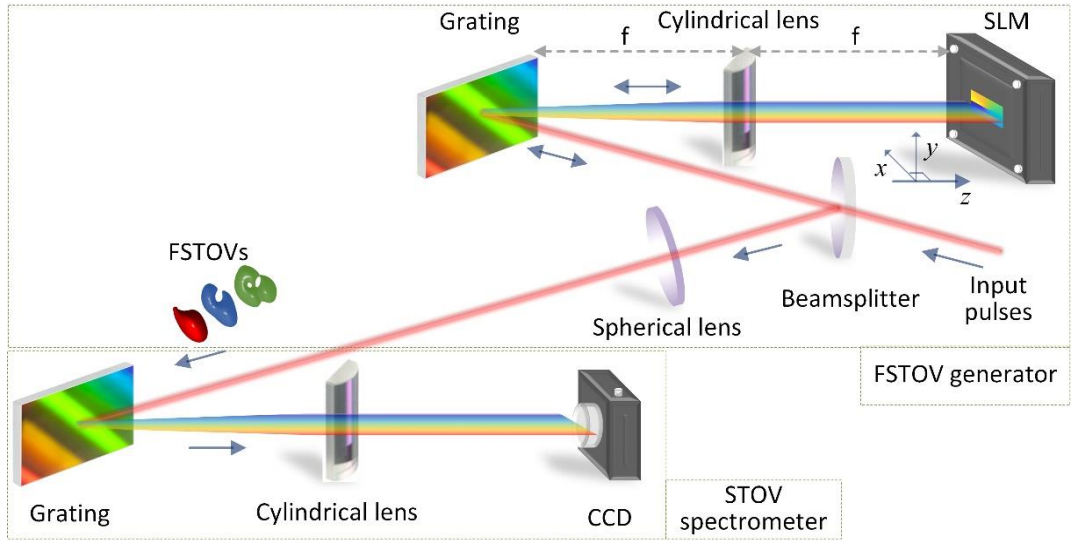

**Fig. 1. Experimental setup for the generation and detection of FSTOV pulses.** The top part shows the FSTOV generator, and the bottom part shows the detection of the diffraction patterns and TCs of the FSTOVs.

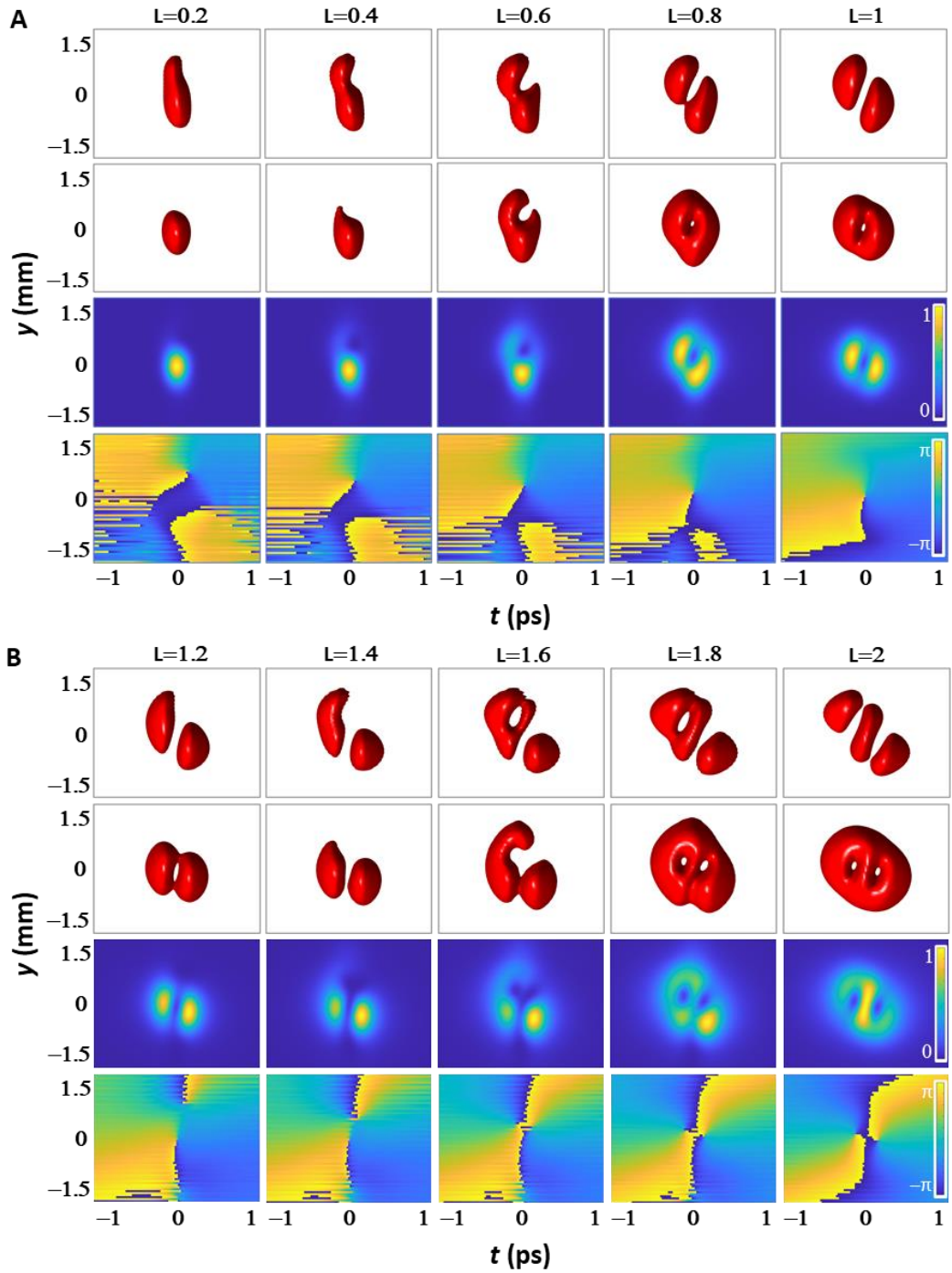

**Fig. 2. Simulated results of the FSTOVs in the near and far fields.** (**A**)Simulated results of the FSTOVs with TCs in the ranges of 0 to 1, and (**B**) 1to 2. In both (**A**) and (**B**), rows 1 and 2 show the 3D-iso-intensity profiles of the FSTOVs in the near and far fields, respectively. Rows 3 and 4 show the intensity profiles and phase patterns of the FSTOVs in the far fields, respectively.

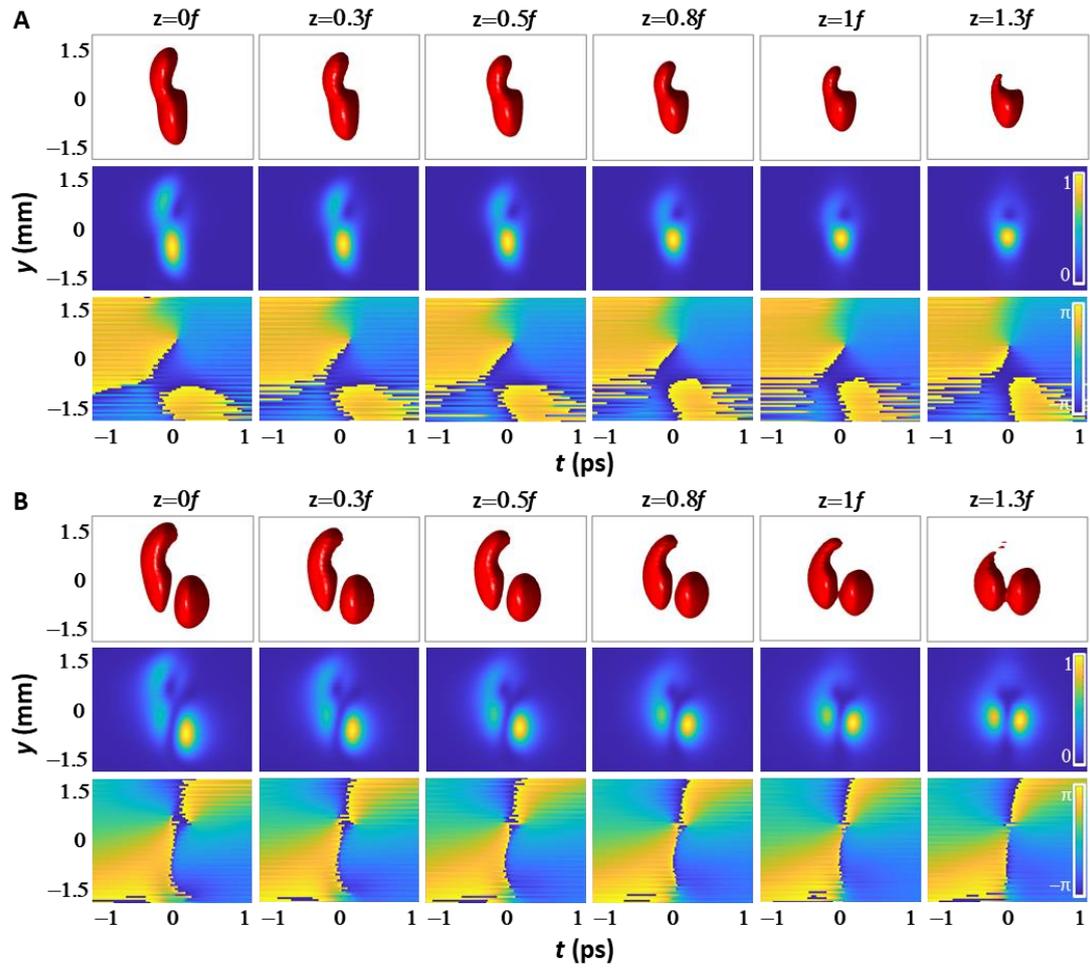

**Fig. 3. The evolution of FSTOVs.** The propagation of FSTOVs with TCs of 0.5 (**A**) and 1 (**B**). In both (**A**) and (**B**), row 1, 2, and 3 show the 3D-isointensity profiles, intensity profiles, and phase patterns of the FSTOVs at different positions after the focal lens.

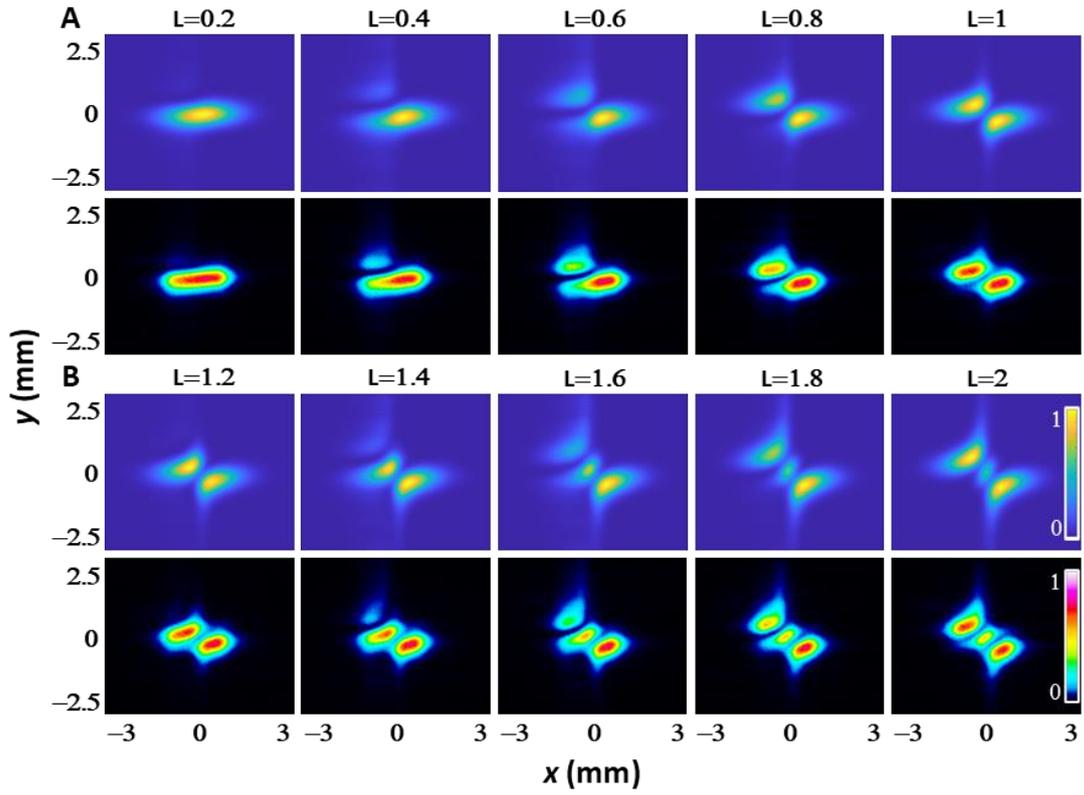

**Fig. 4. Diffraction patterns of the FSTOVs with different TCs.** (**A** and **B**) Diffraction patterns of the FSTOVs with TCs in the ranges of 0 to 1 and 1 to 2, respectively. In (**A** and **B**), the first and second rows show the simulated results and measured results, respectively. (**C** and **D**) The energy ratio of FSTOVs as a function of TC, with TC ranges of 0 to 1 and 1 to 2, respectively.

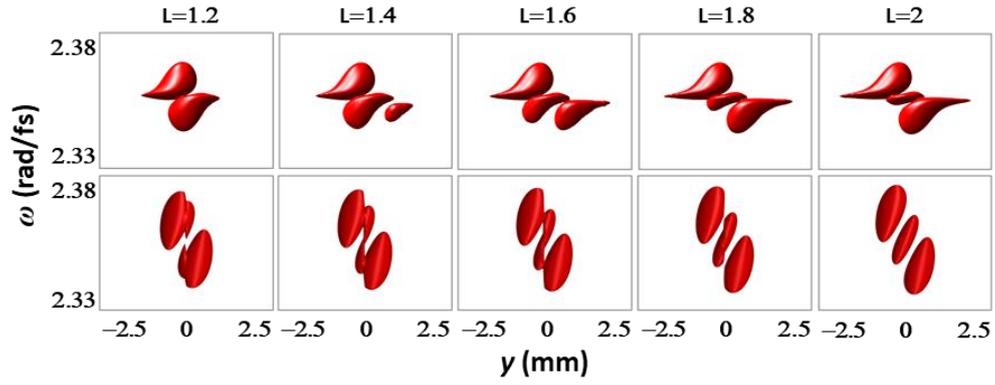

**Fig. 5. Spatial-spectral intensity distributions of the FSTOVs with TCs in the range of 1-2.** Top row shows the spatial-spectral intensity distributions of the FSTOVs generated using the 4*f* pulse shaper. Bottom row shows the spatial-spectral intensity distributions of the FSTOVs calculated using Eq. (1).

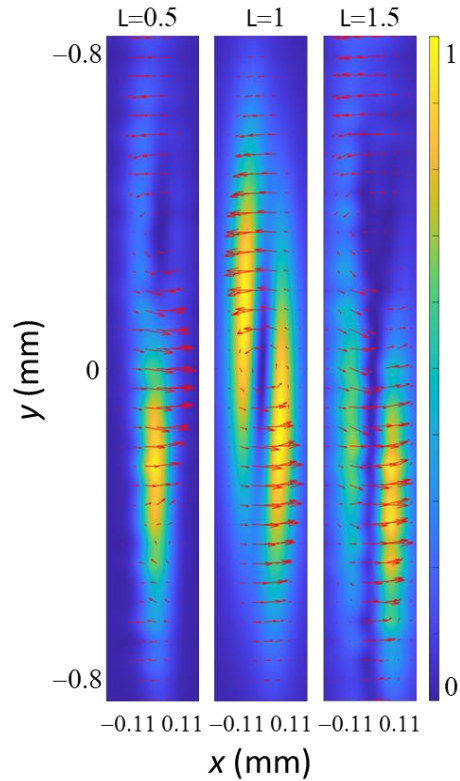

**Fig. 6. The OAM densities of the FSTOVs.** The left, middle and right columns show the OAM densities of the FSTOVs with $l$ = 0.5, 1, and 1.5, respectively.